\documentclass[11pt,a4paper]{article}
\pdfoutput=1

\usepackage{jheppub}

\usepackage{amsmath}
\usepackage{bbm}
\usepackage{float}
\usepackage{graphicx}	
\usepackage{hyperref}
\usepackage{mathtools}
\usepackage{cancel}

\usepackage[normalem]{ulem}

\usepackage{multirow}
\usepackage{rotating}
\usepackage{subcaption}

\def\lsim{\raise0.3ex\hbox{$\;<$\kern-0.75em\raise-1.1ex
\hbox{$\sim\;$}}}
\def\gsim{\raise0.3ex\hbox{$\;>$\kern-0.75em\raise-1.1ex
\hbox{$\sim\;$}}}

\def\thetitle{ 
On the effective atmospheric-scale $\Delta m^2$ in neutrino oscillation \\
 \vspace{- 6mm}
}
\title{\thetitle}
\hypersetup{pdftitle={\thetitle}}
\author{Hisakazu Minakata}
\affiliation{
Tokyo Metropolitan University,\footnote{Professor Emeritus} Minami-Osawa, Hachioji, Tokyo 192-0397, Japan \\ 
}

\emailAdd{hisakazu.minakata@gmail.com}
\date{\today}

\abstract{
We discuss an effective two-frequency description of the three-generation neutrino oscillation with the key quantity $\Delta m^2_{ \text{effective} }$, which 
unifies the effects of atmospheric-scale $\Delta m^2_{31}$-$\Delta m^2_{32}$ driven oscillations. As a solution to the minimization problem in vacuum, Nunokawa {\it et al.} obtained $\Delta m^2_{ \text{effective} } = \Delta m^2_{31} - (2E / L) \arctan \left\{ r_{2} \sin 2\Delta_{21} / ( r_{1} + r_{2} \cos 2\Delta_{21} ) \right\}$, where $\Delta_{21} \equiv \Delta m^2_{21} L / (4 E)$ and $r_{i}$ are the mixing-angle dependent constants. Despite the skepticism for the $L/E$ dependence in this formula, we show, in fact, that this is an exact expression, which renders the probability formula also exact. We offer an appealing-to-intuition physical explanation for the $L/E$ dependence that it originates from the nature-designed three generation neutrinos. The way it is buried into $\Delta m^2_{ \text{effective} }$ is through the solar-effect renormalization to this quantity. With the physical interpretation of $\Delta m^2_{ \text{effective} }$ at hand, we aim at resolving the existing controversies on $\Delta m^2_{ \text{effective} }$ in this paper. } 

\newcommand{\Dmsqren}{\Delta m^2_{ \text{ren} }}

\begin{document} 

\maketitle

\section{Introduction}
\label{sec:introduction} 

As far as the contemporary knowledge teaches us, we live in a world which consists of the three-generations of leptons and quarks~\cite{ParticleDataGroup:2026mpi}. For neutrinos, therefore, we have the three mass-squared differences, $\Delta m^2_{21}$, $\Delta m^2_{31}$, and $\Delta m^2_{32}$, where $\Delta m^2_{ji} \equiv m^2_{j} - m^2_{i}$ ($i, j = 1,2,3$) with $m^2_{j}$ being the neutrino mass squared of $j$-th mass eigenstates. Among them, the only two are independent: $\Delta m^2_{31} = \Delta m^2_{32} + \Delta m^2_{21}$. An important point, which we should all remember, is that this discussion is made possible by the experimental observations of the atmospheric-scale oscillations~\cite{Super-Kamiokande:1998kpq}, and the solar-scale ones~\cite{KamLAND:2002uet,SNO:2002tuh}. The observed values of $\Delta m^2_{ji}$ indicate a hierarchy of a factor of $\sim 30$, that is, $\Delta m^2_{21} / \Delta m^2_{31} \simeq \Delta m^2_{21} / \Delta m^2_{32} \simeq 0.03$. 

Given the hierarchy $\Delta m^2_{31} \simeq \Delta m^2_{32} \gg \Delta m^2_{21}$, it may be natural to seek an effective description of the three-flavor neutrino oscillations with the representative single atmospheric-scale $\Delta m^2$ and the solar-scale $\Delta m^2_{21}$. We denote such unique effective atmospheric-scale $\Delta m^2$ as 
$\Delta m^2_{ \text{effective} }$ in this paper. To our knowledge, the earliest proposal for $\Delta m^2_{ \text{effective} }$ was made by the authors of ref.~\cite{Fogli:2005cq}, $\Delta m^2_{ \text{effective} } = \frac{1}{2} [ \Delta m^2_{31} + \Delta m^2_{32} ]$. Then, the next question may naturally be raised: Is there any way of determining the form of $\Delta m^2_{ \text{effective} }$ by some systematic treatment based on a basic principle? 

In a seminal paper~\cite{Nunokawa:2005nx}, Nunokawa, Parke, and Zukanovich Funchal (NPZ) provided such a formulation to determine the expression of the effective atmospheric $\Delta m^2$ by a suitable optimization procedure. 
They analyzed the disappearance (or, survival) probability in vacuum, $P (\nu_{\alpha} \rightarrow \nu_{\alpha}) = P (\bar{\nu}_{\alpha} \rightarrow \bar{\nu}_{\alpha})$ ($\alpha = e, \mu, \tau$) 
\begin{eqnarray} 
&&
P (\nu_{\alpha} \rightarrow \nu_{\alpha}) 
= 
1 - 4 |U_{\alpha 3}|^2 |U_{\alpha 1}|^2 \sin^2 \Delta_{31} 
- 4 |U_{\alpha 3}|^2 |U_{\alpha 2}|^2 \sin^2 \Delta_{32}
- 4 |U_{\alpha 2}|^2 |U_{\alpha 1}|^2 \sin^2 \Delta_{21},
\nonumber\\
\label{Paa-vac}
\end{eqnarray}
to identify the expression of $\Delta m^2_{ \text{effective} }$. 
In eq.~\eqref{Paa-vac}, $U_{\alpha i}$ denotes the elements of the lepton flavor mixing matrix~\cite{Maki:1962mu}, which connects the neutrino flavor eigenstates to the mass eigenstates as $\nu_{\alpha} = \sum_{i=1,2,3} U_{\alpha i} \nu_{i}$. In this paper, we use (apart from section~\ref{sec:notation}) the form of the $U$ matrix in the Particle Data Group (PDG)~\cite{ParticleDataGroup:2026mpi} convention. 
In eq.~\eqref{Paa-vac} and hereafter, we use the simplified notations 
\begin{eqnarray} 
&&
\Delta_{ji} \equiv \frac{ \Delta m^2_{ji} L }{ 4 E }, 
\label{D-ji-def}
\end{eqnarray}
where $E$ is the neutrino energy and $L$ the baseline distance. 
$\Delta_{31}$ (and $\Delta_{32}$) takes a positive and negative values for 
the normal mass ordering (NMO) and the inverted mass ordering (IMO), respectively.

Via minimization of the solar-scale oscillation terms, NPZ obtained the $\nu_{\alpha}$ survival probability $P (\nu_{\alpha} \rightarrow \nu_{\alpha} )$ in vacuum in the following form~\cite{Nunokawa:2005nx} 
\begin{eqnarray} 
&&
1 - P (\nu_{\alpha} \rightarrow \nu_{\alpha} ) 
= 
4 |U_{\alpha 2}|^2 |U_{\alpha 1}|^2 \sin^2 (\Delta_{21} ) 
\nonumber\\
&+& 
2 |U_{\alpha 3}|^2 ( 1 - |U_{\alpha 3}|^2 ) 
\biggl\{ 1 - \sqrt{ 1 - 4 r_{1} r_{2} \sin^2 \Delta_{21} } 
\cos \left( \frac{ \Delta m^2_{ \text{effective} } L }{ 2E } \right) 
\biggr\}. 
\label{P-aa-NPZ} 
\end{eqnarray}
At the same time, the atmospheric-scale effective $\Delta m^2$, $\Delta m^2_{ \text{effective} }$, is determined by solving the stationary condition as 
\begin{eqnarray} 
&&
\Delta m^2_{ \text{effective} } 
= 
\Delta m^2_{31} - 
\frac{ 2 E }{ L }
\arctan \left\{ \frac{ r_{2} \sin 2\Delta_{21} }{ ( r_{1} + r_{2} \cos 2\Delta_{21} ) } \right\}. 
\label{Dm2-effective-NPZ} 
\end{eqnarray}
$r_{1}$ and $r_{2}$ in eqs.~\eqref{P-aa-NPZ} and~\eqref{Dm2-effective-NPZ} are defined as 
\begin{eqnarray} 
&&
r_{1} \equiv r_{1} (\alpha) = \frac{ |U_{\alpha 1}|^2 }{ |U_{\alpha 1}|^2+|U_{\alpha 2}|^2 }, 
~~~~~~~~~
r_{2} \equiv r_{2} (\alpha) = \frac{ |U_{\alpha 2}|^2 }{ |U_{\alpha 1}|^2+|U_{\alpha 2}|^2 }. 
\label{r1-r2-def}
\end{eqnarray}
The expression of $\Delta m^2_{ \text{effective} }$ in eq.~\eqref{Dm2-effective-NPZ} is a striking result. The solar-scale correction terms in $\Delta m^2_{ \text{effective} }$ obtains the mass dimension {\em not from $\Delta m^2_{21}$} but from the kinematical quantity $E/L$ in this compact form. Furthermore, the additional $E/L$ dependences exist through $\Delta_{21}$ in the argument of arctangent. It means that the effective atmospheric-scale $\Delta m^2_{ \text{effective} }$ is inherently a {\em dynamical quantity}, which varies depending upon $E$ and $L$. A compact recollection of the NPZ minimization procedure will be provided in section~\ref{sec:NPZ-P}. 

Notice that even though we often use the simple notation $r_{i}$ ($i=1,2$) in the various equations suppressing the explicit $\alpha$-dependence, $r_{i} \equiv r_{i} (\alpha)$ do depend on the flavor index $\alpha$ through $| U_{\alpha i} |$. 

In this paper, we address physics of the NPZ formulation and discuss the correctness of $\Delta m^2_{ \text{effective} }$ in eq.~\eqref{Dm2-effective-NPZ}. It appears that people in the community possess the various different opinions on whether $\Delta m^2_{ \text{effective} }$ can be $L/E$ dependent or not. If yes, the question is why is it so?, and what is the physical meaning of the $L/E$ dependence? There might be a confusion among people about which conditions must be satisfied by 
$\Delta m^2_{ \text{effective} }$. In this article, we try to make the concept of $\Delta m^2_{ \text{effective} }$ crystal-clear, and attempt at resolving controversies, if any, on the atmospheric-scale $\Delta m^2$. As no candidate principle is known, besides the NPZ procedure, to obtain $\Delta m^2_{ \text{effective} }$ we concentrate on the approach along this line of thought in this paper. 

There exist the three key statements in this paper: 
\begin{itemize}

\item 
The NPZ probability eq.~\eqref{P-aa-NPZ} is an exact expression, so is $\Delta m^2_{ \text{effective} }$. 

\item 
$L/E$ dependence in $\Delta m^2_{ \text{effective} }$ is what nature chooses.  

\item 
There exists certain kinematical region where $\Delta m^2_{ \text{effective} }$ behaves as a set of the mass and mixing parameters in a good approximation. But, this constant appear to be neither fundamental, nor a building block of the theory of neutrino oscillation. 

\end{itemize}

\noindent
It appears that the first statement above is known~\cite{SP-private}. However, it has not been published in the literature to the best of the author's knowledge. 

The paper plan: 
In sections~\ref{sec:MNPZ-P},~\ref{sec:NPZ-P},~and \ref{sec:equivalence}, we present the key mathematical formulas for the oscillation probability and $\Delta m^2_{ \text{effective} }$, which will be fully utilized in the physics discussions in the rest of this paper. In sections~\ref{sec:L-E-dep-meaning},~\ref{sec:Daya-Bay}, and~\ref{sec:notation}, we discuss the meaning of the $L/E$ dependence in $\Delta m^2_{ \text{effective} }$, and make the concept of $\Delta m^2_{ \text{effective} }$ fully transparent. 
Throughout this paper, we restrict our discussions in vacuum environment, apart from that in section~\ref{sec:Dm2-eff-matter} where we give a brief treatment in matter. In section~\ref{sec:conclusion}, we summarize our results and conclude with some remarks. 

\section{Alternative from of an exact vacuum oscillation probability}
\label{sec:MNPZ-P}

The most transparent way of presenting our discussion in this paper is to start from an alternative form of the vacuum oscillation probability given in eq.~\eqref{Paa-vac}. For brevity, we denote this new form of the probability as the MNPZ expression~\cite{Minakata:2007tn}. The MNPZ form of the survival $\nu_{\alpha}$ probability in vacuum reads: 
\begin{eqnarray} 
&&
1 - P (\nu_{\alpha} \rightarrow \nu_{\alpha}) 
\nonumber\\
&=& 
2 |U_{\alpha 3}|^2 ( 1 -  |U_{\alpha 3}|^2 ) 
\biggl\{ 1 - \sqrt{ 1 - 4 r_{1} r_{2} \sin^2 \Delta_{21} } 
\cos \left( 2 \Delta_{\alpha \alpha} \pm \phi_{\odot} \right) 
\biggr\} 
+ 4 |U_{\alpha 2}|^2 |U_{\alpha 1}|^2 \sin^2 (\Delta_{21} ). 
\nonumber\\
\label{P-aa-MNPZ} 
\end{eqnarray}
In eq.~\eqref{P-aa-MNPZ} the upper (+) and lower (-) signs at $\phi_{\odot}$ imply the NMO and IMO, respectively. 
$\Delta_{\alpha \alpha}$ is defined by 
\begin{eqnarray} 
&&
\Delta_{\alpha \alpha} 
= 
r_{1} (\alpha) | \Delta_{31} | + r_{2} (\alpha) | \Delta_{32} | > 0. 
\label{Daa-def-alpha} 
\end{eqnarray}
The angle $\phi_{\odot}$ is defined by 
\begin{eqnarray} 
&&
\cos \phi_{\odot} 
= 
\frac{ r_{1} \cos ( 2 r_{2} \Delta_{21} ) + r_{2} \cos ( 2 r_{1} \Delta_{21} ) }
{ \sqrt{ 1 - 4 r_{1} r_{2} \sin^2 \Delta_{21} }  }, 
\nonumber\\
&& 
\sin \phi_{\odot} 
= 
\frac{ r_{1} \sin ( 2 r_{2} \Delta_{21} ) - r_{2} \sin ( 2 r_{1} \Delta_{21} ) }
{ \sqrt{ 1 - 4 r_{1} r_{2} \sin^2 \Delta_{21} }  }. 
\label{sin-cos-phi-expression} 
\end{eqnarray}
We remark that $r_{i} \equiv r_{i} (\alpha)$ ($i=1,2$) in eqs.~\eqref{P-aa-MNPZ} and~\eqref{sin-cos-phi-expression}, the notation prevails throughout this paper. 
Alternatively to eq.~\eqref{sin-cos-phi-expression}, one can write 
\begin{eqnarray} 
&&
\phi_{\odot} 
= 
\arctan \left\{ ( r_{1} - r_{2} ) \tan \Delta_{21} \right\} 
- ( r_{1} - r_{2} ) \Delta_{21}. 
\label{phi-expression}
\end{eqnarray}
A brief account of the equality between $\phi_{\odot}$ in eqs.~\eqref{sin-cos-phi-expression} and~\eqref{phi-expression} will be given in Appendix~\ref{sec:phi-relations}. 

By using eqs.~\eqref{Daa-def-alpha} and~\eqref{sin-cos-phi-expression}, one can easily show that 
\begin{eqnarray} 
&&
\cos \left( 2 \Delta_{\alpha \alpha} \pm \phi_{\odot} \right) 
= 
\frac{1}{ \sqrt{ 1 - 4 r_{1} r_{2} \sin^2 \Delta_{21} }  } 
\biggl\{ r_{1} \cos ( 2 \Delta_{31} ) + r_{2} \cos ( 2 \Delta_{32} ) \biggr\} 
\label{inter-step}
\end{eqnarray} 
By inserting eq.~\eqref{inter-step} into eq.~\eqref{P-aa-MNPZ} one reproduces the original expression of the vacuum oscillation probability in eq.~\eqref{Paa-vac}.\footnote{
In the $\nu_{e}$ channel, $r_{1} = c^2_{12}$, $r_{2} = s^2_{12}$, and $4 r_{1} r_{2} = \sin^2 2\theta_{12}$. }
Therefore, the MNPZ form of the probability eq.~\eqref{P-aa-MNPZ} is equivalent to eq.~\eqref{Paa-vac}, and it is exact. 
The MNPZ probability will play a key role in our discussions in this paper. See sections~\ref{sec:equivalence} and~\ref{sec:L-E-dep-meaning}. For convenience, we often refer the argument of cosine in eqs.~\eqref{P-aa-NPZ} and~\eqref{P-aa-MNPZ} as the ``oscillation factor'' in this paper. 

The authors of ref.~\cite{Minakata:2007tn} make an extensive use of the probability eq.~\eqref{P-aa-MNPZ} to present the transparent discussion of the effect of neutrino mass orderings, an advanced (NMO) versus retarded (IMO) waves due to the phase $\phi_{\odot}$, as one can observe in eq.~\eqref{P-aa-MNPZ}. The notation $\phi_{\odot}$ indicates that it is governed only by the solar-scale $\Delta_{21}$, which separates its role from the atmospheric-scale $\Delta_{\alpha \alpha}$. 
The probability formula in eq.~\eqref{P-aa-MNPZ} with the matter-effect corrections is used to estimate the sensitivity to the mass-ordering determination in a JUNO-only setting, which results in slightly below $2 \sigma$~\cite{Capozzi:2013psa}. 

\section{NPZ probability and $\Delta m^2_{ \text{effective} }$}
\label{sec:NPZ-P}

In ref.~\cite{Nunokawa:2005nx}, NPZ ask a question: Is it possible to find the concrete expression of the effective atmospheric $\Delta m^2$ by a certain suitable optimization procedure? Remarkably, they found an affirmative answer. The rest of this section is a compact recollection of the NPZ treatment. 

\subsection{Derivation of the NPZ probability}
\label{sec:NPZ-derivation}

They define a candidate expression of the effective atmospheric $\Delta m^2$ as follows:
\begin{eqnarray} 
&&
\Delta_{\eta} \equiv \Delta_{31} - \eta \Delta_{21}, 
\label{eta-def}
\end{eqnarray}
where $\Delta_{31} \equiv \Delta m^2_{31} L / (4E)$, etc. as defined in eq.~\eqref{D-ji-def}. Then, they show that the vacuum oscillation probability in eq.~\eqref{Paa-vac} can be written as 
\begin{eqnarray} 
&&
1 - P (\nu_{\alpha} \rightarrow \nu_{\alpha})
\nonumber\\
&=& 
4 |U_{\alpha 3}|^2 ( 1 - |U_{\alpha 3}|^2 ) 
\biggl[ 
\sin^2 \Delta_{\eta }
+ \left\{ r_{1} \sin^2 (\eta \Delta_{21} ) + r_{2} \sin^2 [ ( 1 - \eta ) \Delta_{21} ] 
\right\} 
\cdot 
\cos 2\Delta_{\eta} 
\nonumber\\
&+& 
\frac{d}{ d ( \eta \Delta_{21} ) } 
\left\{ r_{1} \sin^2 ( \eta \Delta_{21} ) + r_{2} \sin^2 [ ( 1 - \eta ) \Delta_{21} ] \right\} 
\cdot 
\sin 2\Delta_{\eta} 
\biggr] 
+ 4 |U_{\alpha 2}|^2 |U_{\alpha 1}|^2 \sin^2 (\Delta_{21} ). 
\nonumber \\
\label{NPZ-1}
\end{eqnarray}
NPZ propose to determine $\eta$ by minimizing the two small solar-scale terms, the ones with the curly bracket in eq.~\eqref{NPZ-1}. If we choose $\eta$ so as to set the coefficient in front of $\sin 2\Delta_{\eta}$ to vanish, 
\begin{eqnarray} 
&&
\frac{d}{ d ( \eta \Delta_{21} ) } 
\left\{ r_{1} \sin^2 ( \eta \Delta_{21} ) + r_{2} \sin^2 [ ( 1 - \eta ) \Delta_{21} ] \right\} 
= 
r_{1} \sin ( 2 \eta \Delta_{21} ) 
- r_{2} \sin [ 2 ( 1 - \eta ) \Delta_{21} ] = 0, 
\nonumber \\
\label{eta-determine}
\end{eqnarray}
$\left\{ r_{1} \sin^2 ( \eta \Delta_{21} ) + r_{2} \sin^2 [ ( 1 - \eta ) \Delta_{21} ] \right\}$, the coefficient of $\cos 2\Delta_{\eta}$ takes a extrema, either a minimum or a maximum at this value of $\eta$, which we denote $\eta_{0}$. One can show easily that it is a minimum. 

Using eq.~\eqref{eta-determine}, $\eta_{0}$ can be determined as 
\begin{eqnarray} 
&&
\eta_{0} 
= 
\frac{1}{ 2 \Delta_{21} } 
\arctan \left\{ \frac{ r_{2} \sin 2\Delta_{21} }{ ( r_{1} + r_{2} \cos 2\Delta_{21} ) } \right\}. 
\label{eta-solution}
\end{eqnarray}
Then, the effective atmospheric $\Delta m^2$ is determined as $\Delta_{\eta=\eta_{0}} = \Delta m^2_{ \text{effective} } L / (4E)$. The explicit expression of $\Delta m^2_{ \text{effective} }$ is given by eq.~\eqref{Dm2-effective-NPZ}. Finally, an identity 
\begin{eqnarray} 
&&
\arctan x = \arcsin \frac{ x }{ \sqrt{ 1 + x^2 } } = \arccos \frac{ 1 }{ \sqrt{ 1 + x^2 } }
\label{identity-1} 
\end{eqnarray}
can be used to derive the formula at $\eta = \eta_{0}$: 
\begin{eqnarray} 
&&
r_{1} \sin^2 (\eta_{0} \Delta_{21} ) + r_{2} \sin^2 [ ( 1 - \eta_{0} ) \Delta_{21} ] 
=  
\frac{ 1 }{2} 
\biggl\{ 1 - \sqrt{ 1 - 4 r_{1} r_{2} \sin^2 \Delta_{21} } \biggr\}. 
\label{formula-1}
\end{eqnarray} 
By collecting the formulas in eqs.~\eqref{eta-def},~\eqref{eta-solution}, and~\eqref{formula-1}, the NPZ form of the survival probability eq.~\eqref{P-aa-NPZ}, and the expression of $\Delta m^2_{ \text{effective} }$ in eq.~\eqref{Dm2-effective-NPZ} can readily be derived. 

\subsection{Nearly $L/E$ independent $\Delta m^2_{ \text{effective} }$ in region $\Delta_{21} \ll 1$}
\label{sec:constant-Dm2} 

One may ask a question: Is there any kinematical region where $L/E$ independence of $\Delta m^2_{ \text{effective} }$ is approximately realized. If we expand the arctangent term in eq.~\eqref{Dm2-effective-NPZ} in terms of $\Delta m_{21}$, the leading term is $L/E$ independent as $\arctan x \approx x$ at $x \ll 1$. Then, we expand $\Delta m^2_{ \text{effective} }$ in power series of $\Delta_{21}$. We obtain, to order $\mathcal{O} [ (\Delta_{21})^7 ]$, 
\begin{eqnarray} 
&&
\Delta m^2_{ \text{effective} } 
= 
\Delta m^2_{31} - r_{2} \Delta m^2_{21} 
\nonumber\\
&-& 
r_{2} \Delta m^2_{21} 
\biggl[ 
\frac{ 2 }{ 3 } r_{1} ( r_{2} - r_{1} ) ( \Delta_{21} )^2 
+ \left\{ \frac{ 2 }{ 15 } - ( 2 r_{2} ) + \frac{ 5 }{ 3 } ( 2 r_{2} )^2 - ( 2 r_{2} )^3 
+ \frac{1}{5} ( 2 r_{2} )^4 \right\} 
( \Delta_{21} )^4 
\nonumber\\
&+& 
\left\{ - \frac{ 4 }{ 315 } + \frac{ 28 }{ 45 } ( 2 r_{2} ) 
- \frac{ 86 }{ 45 } ( 2 r_{2} )^2 + \frac{ 10 }{ 3 } ( 2 r_{2} )^3 
- \frac{ 8 }{ 3 } ( 2 r_{2} )^4 + ( 2 r_{2} )^5 \right\} 
( \Delta_{21} )^6 
+ 
\mathcal{O} \left\{ ( \Delta_{21} )^8 \right\} 
\biggr]. 
\nonumber\\
\label{Dm2-expanded}
\end{eqnarray}
The leading, solar-scale $- r_{2} \Delta m^2_{21}$ term in the first line is the anticipated unique $L/E$ independent contribution from the arctangent term in eq.~\eqref{Dm2-effective-NPZ}. The rest of the terms are all $L/E$ dependent and involve the terms of the order of $( \Delta_{21} )^3$, or higher.\footnote{
In section~\ref{sec:introduction}, we have stated that: ``the solar-scale correction terms in $\Delta m^2_{ \text{effective} }$ obtains the right dimension {\em not from $\Delta m^2_{21}$} but from the kinematical quantity $E/L$.'' It is interesting to note that eq.~\eqref{Dm2-expanded} provides an alternative view that the solar-scale correction terms in $\Delta m^2_{ \text{effective} }$ obtains the right dimension from $\Delta m^2_{21}$. Each statement is true depending upon one's choice of the expression of $\Delta m^2_{ \text{effective} }$, the compact form as in eq.~\eqref{Dm2-effective-NPZ}, or the perturbative form as in eq.~\eqref{Dm2-expanded}. Anyway, what is certainly true in the both cases is that $\Delta m^2_{ \text{effective} }$ has the profound $L/E$ dependences. }

In region where $( \Delta_{21} ) \ll 1$, which may be justified in the short-baseline reactor neutrino experiments, e.g., for the $\theta_{13}$ measurement~\cite{DayaBay:2018yms,RENO:2020dxd,DoubleChooz:2019qbj}, 
\begin{eqnarray} 
&& 
\Delta_{21} \equiv 
\frac{ \Delta m^2_{21} L}{4 E} 
= 
3.18 \times 10^{-2} 
\left(\frac{ \Delta m^2_{21} }{7.5 \times 10^{-5}\mbox{eV}^2}\right) 
\left(\frac{L}{1 \mbox{km}}\right)
\left(\frac{E}{3 \mbox{MeV}}\right)^{-1}, 
\label{solar-Delta}
\end{eqnarray}
we can use the first line in eq.~\eqref{Dm2-expanded} as an approximate $\Delta m^2_{ \text{effective} }$. 
This is nothing but the $L/E$ independent $\Delta m^2_{\alpha \alpha}$ defined by 
\begin{eqnarray} 
&&
\Delta m^2_{\alpha \alpha} = \Delta m^2_{31} - r_{2} \Delta m^2_{21}, 
\label{Dm2aa-def}
\end{eqnarray} 
which is advocated by NPZ for phenomenological analyses. The term ignored in eq.~\eqref{Dm2aa-def} is a factor of $\simeq 3 \times 10^{-5}$ smaller compared to the leading $\Delta m^2_{31}$ term. 
Equation~\eqref{Dm2aa-def} is consistent with eq.~\eqref{Daa-def-alpha}. There must be no confusion between the two different objects, ($\alpha$-dependent) $\Delta m^2_{ \text{effective} }$ as the effective atmospheric $\Delta m^2$ given by 
eq.~\eqref{Dm2-effective-NPZ}, and their $L/E$ independent part $\Delta m^2_{\alpha \alpha}$. 

Use of $\Delta m^2_{\alpha \alpha}$ in the contexts of phenomenological analyses is extensively discussed, see e.g., ref.~\cite{Parke:2016joa}. While we do not go into details we mention that the utility of $\Delta m^2_{\alpha \alpha}$ can become manifest in the simple formula for the mass ordering determination, $\Delta m^2_{ \text{ee} } - \Delta m^2_{\mu \mu}$ is positive (negative) for the NMO (IMO)~\cite{Nunokawa:2005nx}, which triggered further examinations. See, for example, refs.~\cite{Minakata:2006gq,Parke:2016joa,Parke:2024xre,Alves:2026ydc}.  
The coming results from the ongoing and future precision measurements in the $\nu_{e}$ and $\nu_{\mu}$ disappearance channels~\cite{JUNO:2015zny,Hyper-Kamiokande:2018ofw,DUNE:2020jqi} would allow a decisive statement on the neutrino mass ordering. 

\section{Equivalence between the NPZ and MNPZ probabilities} 
\label{sec:equivalence} 

We notice that the forms of the NPZ probability in eq.~\eqref{P-aa-NPZ} and the MNPZ probability in eq.~\eqref{P-aa-MNPZ} are so akin to each other. Then, a natural ansatz which deserves a close examination is that they are identical. The identity holds if 
\begin{eqnarray} 
&&
2 \Delta_{\alpha \alpha} + \phi_{\odot} = \frac{ \Delta m^2_{ \text{effective} } L }{ 2E }
\label{exactness-condition} 
\end{eqnarray}
for arbitrary $E$ and $L$. In this section we will prove that eq.~\eqref{exactness-condition} holds. 

Using eq.~\eqref{phi-expression}, the left-hand side of eq.~\eqref{exactness-condition} is given by 
\begin{eqnarray} 
&&
2 \Delta_{\alpha \alpha} + \phi_{\odot} 
=  
2 \Delta_{31} + \arctan \left( ( r_{1} - r_{2} ) \tan \Delta_{21} \right) - \Delta_{21}. 
\label{LHS}
\end{eqnarray}
Using eq.~\eqref{Dm2-effective-NPZ}, the [left-hand side - right-hand side] of eq.~\eqref{exactness-condition} can be written as 
\begin{eqnarray} 
&&
\hspace{-4mm}
2 \Delta_{\alpha \alpha} + \phi_{\odot} - \frac{ \Delta m^2_{ \text{effective} } L }{ 2E } 
= 
\arctan \left\{ ( r_{1} - r_{2} ) \tan \Delta_{21} \right\} - \Delta_{21} 
+ \arctan \left\{ \frac{ r_{2} \sin 2\Delta_{21} }{ ( r_{1} + r_{2} \cos 2\Delta_{21} ) } \right\}. 
\nonumber\\
\label{LHS-RHS} 
\end{eqnarray}
To show that the right-hand side of eq.~\eqref{LHS-RHS} vanishes we use a formula, 
\begin{eqnarray} 
&&
\arctan x \pm \arctan y 
= 
\arctan \frac{ x \pm y }{ 1 \mp x y }. 
\label{math-1}
\end{eqnarray}
Taking the upper sign and inserting $x= \left\{ ( r_{1} - r_{2} ) \tan \Delta_{21} \right\}$ and $y = \left\{ \frac{r_2 \sin 2\Delta_{21}}{r_1+r_2\cos2\Delta_{21}} \right\}$ into eq.~\eqref{math-1}, we obtain 
\begin{eqnarray} 
&&
\arctan \left\{ ( r_{1} - r_{2} ) \tan \Delta_{21} \right\} 
+ \arctan \left\{ \frac{r_2 \sin 2\Delta_{21}}{r_1+r_2\cos2\Delta_{21}} \right\} 
= 
\Delta_{21} 
\label{identity}
\end{eqnarray}
This implies that the right-hand side of eq.~\eqref{LHS-RHS} vanishes and the equivalence condition eq.~\eqref{exactness-condition} holds.  

Thus, we have proven that the NPZ probability eq.~\eqref{P-aa-NPZ} and the MNPZ probability eq.~\eqref{P-aa-MNPZ} are identical to each other. By construction, the MNPZ probability is the same as the original oscillation probability in eq.~\eqref{Paa-vac}, and therefore the both NPZ and MNPZ expressions give the exact expression of the vacuum oscillation probability $P (\nu_{\alpha} \rightarrow \nu_{\alpha})$ and $P (\bar{\nu}_{\alpha} \rightarrow \bar{\nu}_{\alpha})$.

\subsection{Intermediate summary and what is left for the exact $\Delta m^2_{ \text{effective} }$}
\label{sec:exactness} 

In section~\ref{sec:introduction}, we have introduced the problem of obtaining the single atmospheric-scale effective $\Delta m^2$ in seeking for the two-frequency description of the three-flavor neutrino oscillations. In section~\ref{sec:NPZ-P}, by minimizing the solar-scale terms, we have obtained the expressions of the probability and $\Delta m^2_{ \text{effective} }$~\`a la NPZ~\cite{Nunokawa:2005nx}. 
It turned out that the resultant expression of $P(\nu_{\alpha} \rightarrow \nu_{\alpha})$ in eq.~\eqref{P-aa-NPZ} with this $\Delta m^2_{ \text{effective} }$ is identical with the exact oscillation probability. 
A proof goes through via the two steps: (i) the exact probability is rewritten in the MNPZ form, eq.~\eqref{P-aa-MNPZ}, and (ii) we show the equivalence between the MNPZ and the NPZ forms. 

Therefore, the problem of obtaining the unique form of the effective atmospheric-scale $\Delta m^2$ is solved by $\Delta m^2_{ \text{effective} }$ given in eq.~\eqref{Dm2-effective-NPZ}, as far as the disappearance channels in vacuum is concerned. It has settled all possible controversies between the various candidate forms of $\Delta m^2_{ \text{effective} }$, because the NPZ form provides the exact solution. 

While this observation, in principle, must bring discussions on ``what is the effective atmospheric-scale $\Delta m^2$'' to an end,  some critiques, or reservations may still remain: 
\begin{itemize}

\item 
$\Delta m^2_{ \text{effective} }$ in eq.~\eqref{Dm2-effective-NPZ} is $L/E$ dependent, which is unphysical for the effective $\Delta m^2$. The whole concept of the effective atmospheric $\Delta m^2$ is under suspicion. 

\item 
One may propose to recognize only the $L/E$ independent part of $\Delta m^2_{ \text{effective} }$, eq.~\eqref{Dm2-expanded}, as the effective atmospheric $\Delta m^2$. 

\end{itemize}
Despite skepticism on the effective atmospheric $\Delta m^2$ stated above, in the rest of this paper, we try to defend the concept of $\Delta m^2_{ \text{effective} }$. We address the conceptual issues on $\Delta m^2_{ \text{effective} }$ in eq.~\eqref{Dm2-effective-NPZ} on which the skepticisms are based, and present the correct interpretation of it in the light of our newly gained aspects, the exactness of $\Delta m^2_{ \text{effective} }$. After all, we will see that the key to understand the $\Delta m^2_{ \text{effective} }$ is provided by the nature created design of our world with the three-generation neutrinos. 

\section{Meaning of the $L/E$ dependence in $\Delta m^2_{ \text{effective} }$}
\label{sec:L-E-dep-meaning} 

What is the meaning of the $L/E$ dependence in $\Delta m^2_{ \text{effective} }$? 
This is the most important question for the purpose of this paper. Even though the mass dimension of the solar-scale correction terms in $\Delta m^2_{ \text{effective} }$ may be provided either by $L/E$ or $\Delta m^2_{21}$ in the compact (eq.~\eqref{Dm2-effective-NPZ}) or perturbative (eq.~\eqref{Dm2-expanded}) expressions, respectively, existence of the profound $L/E$ dependences in $\Delta m^2_{ \text{effective} }$ is not disputable. Then, what does it mean? 

\subsection{$\Delta m^2_{ \text{effective} }$ has to be $L/E$ dependent} 
\label{sec:L-E-dep-must} 

We can provide a simple argument that if $\Delta m^2_{ \text{effective} }$ lacks $L/E$ dependence, it cannot describe the generic features of the three-generation neutrino oscillations. In such system, we have the three different oscillation frequencies associated with $\Delta m^2_{31}$, $\Delta m^2_{32}$, and $\Delta m^2_{21}$, barring degeneracies. Without $L/E$ dependence in $\Delta m^2_{ \text{effective} }$ the effective description of the system has only the two different constant frequencies, the ones associated with $\Delta m^2_{ \text{effective} }$ and $\Delta m^2_{21}$. Clearly, such system cannot describe the three-generation neutrino oscillations. 

Notice that the fact that the only two $\Delta m^2$ out of three are independent cannot help. In fact, this point is irrelevant for our present discussion even though it might have caused a confusion. Without degeneracy in neutrino masses, the three neutrino oscillation phenomena contain the three different frequencies which can be measured, in principle, by the experiments. This system cannot be mimicked by the two constant-frequency system. 

\subsection{Exceptional case with approximate mass degeneracy} 
\label{sec:small-D21} 

The above discussions need modification if a degeneracy exists in the neutrino mass spectrum, $m_{2} = m_{1}$, for example. This system has the unique mass-squared difference $\Delta_{31} = \Delta_{32}$, and consequently only the atmospheric-scale oscillation occurs. 
If the degeneracy is approximate, $\Delta_{21} \ll \Delta_{31} \approx \Delta_{32}$, another exceptional situation emerges. At around the atmospheric oscillation maximum $\Delta_{31} \approx \Delta_{32} \simeq 1$, and $\Delta_{21} \simeq  \Delta m^2_{21} / \Delta m^2_{31} \ll 1$. Then, $\Delta m^2_{ \text{effective} }$ can be expanded in terms of $\Delta_{21}$ as done in section~\ref{sec:constant-Dm2}, see eq.~\eqref{Dm2-expanded}. If we keep the terms up to order $\Delta_{21}$, we obtain $\Delta m^2_{ \text{effective} } \approx  \Delta m^2_{\alpha \alpha}$, where $\Delta m^2_{\alpha \alpha}$ is the $L/E$-independent part of $\Delta m^2_{ \text{effective} }$ defined in eq.~\eqref{Dm2aa-def}. The ignored third-order term in $\Delta_{21}$ is of the order of $3 \times 10^{-5}$, see eq.~\eqref{solar-Delta}. 
Therefore, the $L/E$-independent approximate formula $\Delta m^2_{ \text{effective} } = \Delta m^2_{\alpha \alpha}$ may be used in analyzing the short-baseline reactor antineutrino experiments for $\theta_{13}$, Daya Bay~\cite{DayaBay:2018yms}, RENO~\cite{RENO:2020dxd}, and Double-Chooz~\cite{DoubleChooz:2019qbj}. 

\subsection{An inverse problem and the raison d'\^etre of the $L/E$ dependence }
\label{sec:inverse-problem} 

To have better understanding of the $L/E$-dependence in $\Delta m^2_{ \text{effective} }$ along the reasoning in section \ref{sec:L-E-dep-must}, let us ask a pedagogical question: Why is introduction of appropriate $L/E$ dependence in $\Delta m^2_{ \text{effective} }$ sufficient to describe {\em exactly} the three-generation world? In the above we have made a convincing case that the $\Delta m^2_{ \text{effective} }$ - $\Delta m^2_{21}$ two constant-frequency system cannot describe the three-generation neutrino oscillation which is equipped with the three different frequencies. Then, why and how does just letting one of the two frequencies $L/E$-dependent make it possible? 

We know the answer to this question mathematically. Existence of the exact NPZ or MNPZ probabilities, see eqs.~\eqref{P-aa-NPZ} and~\eqref{P-aa-MNPZ}, proves that the $L/E$-dependent $\Delta m^2_{ \text{effective} }$ is sufficient to describe the three-generation neutrino oscillation. 
If we learn how the $L/E$-dependence is buried into $\Delta m^2_{ \text{effective} }$ in these formulas, it should illuminate why the $L/E$ dependence must exist in $\Delta m^2_{ \text{effective} }$, and in which way it does. While this may not tell directly why the $L/E$ dependence provides us the sufficient condition for the exactness, it will reveal the physical reason for the $L/E$ dependence. 

Then, let us examine the MNPZ probability~eq.~\eqref{P-aa-MNPZ}, in which the oscillation factor in cosine is separated into the two parts, $\Delta m^2_{ \text{effective} } L / (2E) = 2 \Delta_{\alpha \alpha} \pm \phi_{\odot}$. This is the decomposition into a completely different two entities, (1) atmospheric-scale $L/E$ independent $\Delta_{\alpha \alpha}$ in eq.~\eqref{Daa-def-alpha} (or eq.~\eqref{Dm2aa-def}), and (2) the solar-scale $L/E$ dependent $\phi_{\odot}$ in eq.~\eqref{sin-cos-phi-expression}. 
Due to the disparity in size of the two $\Delta m^2$, the solar oscillation has a different $L/E$ dependence from the atmospheric one, letting $\Delta_{\alpha \alpha}$ and $\phi_{\odot}$ in the oscillation factor having a different $L/E$ dependence to each other's. For $L/E$-independent $\Delta_{\alpha \alpha}$, $\phi_{\odot}$ must have the appropriate $L/E$ dependence, giving a natural reason for the $L/E$ dependent $\phi_{\odot}$ in $\Delta m^2_{ \text{effective} }$. 

To summarize, we observe that: 
\begin{itemize}

\item 
The effective two-frequency description of the three-neutrino oscillation requires the atmospheric-scale $\Delta m^2_{ \text{effective} }$ which is renormalized by the effect of solar oscillation. 

\item 
Coexistence of the atmospheric and solar frequencies in the oscillation factor is the very reason for $L/E$-dependence of $\Delta m^2_{ \text{effective} }$. 

\end{itemize}

\subsection{Order unity $\Delta_{21}$ region requires picture changing}
\label{sec:picture-change} 

In region where $\Delta_{21} \sim \mathcal{O} (1)$, all the higher order terms of $\Delta_{21}$ with $L/E$ dependence in eq.~\eqref{Dm2-expanded} have the same order of magnitude with the leading, order $\Delta_{21}$ correction term. Then, of course, there is no reason to ignore these higher order terms. Therefore, we need a new picture for  intuitive understanding the $\Delta_{21} \sim \mathcal{O} (1)$ region. 

A small reorganization of the oscillation factor in cosine in \eqref{P-aa-NPZ}  brings us a picture changing which is appropriate to describe the $\Delta_{21} \sim \mathcal{O} (1)$ region. Just pick up the above mentioned order $\Delta_{21}$ term from $\Delta m^2_{ \text{effective} }$ in eq.~\eqref{Dm2-effective-NPZ}, and move it to the first line as done in eq.~\eqref{Dm2-expanded}. This is nothing but the decomposition $\Delta m^2_{ \text{effective} } L / (2E) = 2 \Delta_{\alpha \alpha} \pm \phi_{\odot}$, or the identity eq.~\eqref{exactness-condition}, which is equipped in the MNPZ probability formula. 
In section~\ref{sec:inverse-problem} we have emphasized that this reorganization makes the character of $\Delta m^2_{ \text{effective} }$ illuminating. It decomposes the oscillation factor into (1) the atmospheric-scale constant $\Delta_{\alpha \alpha}$, and (2) the solar-scale $L/E$ dependent $\phi_{\odot}$ in eq.~\eqref{phi-expression}. We can see the latter part as an ``organized'' solar-scale oscillation phase $\phi_{\odot}$, instead of collection of many ``dumb'' terms with $L/E$ dependence. 

A new picture emerged from reorganization of the oscillation factor: In region $\Delta_{21} \sim \mathcal{O} (1)$, the three-generation neutrino oscillation organizes itself as the large-amplitude solar-scale slow oscillations, and the fast atmospheric oscillations (wiggles) superimposed on it. This is a intuition-appealing new picture in the $\Delta_{21} \sim \mathcal{O} (1)$ region emphasized in ref.~\cite{Minakata:2007tn}. 

\section{Use of $\Delta m^2_{ \text{effective} }$ in the Daya Bay analysis} 
\label{sec:Daya-Bay}

It is likely that the short-baseline reactor antineutrino experiments are the best place for the $\Delta m^2_{ \text{effective} }$ description of the disappearance probability $P (\bar{\nu}_{e} \rightarrow \bar{\nu}_{e}) = P (\nu_{e} \rightarrow \nu_{e})$. Naturally, we suggest to utilize the survival probability in eq.~\eqref{P-aa-NPZ} with $\Delta m^2_{ \text{effective} }$ in eq.~\eqref{Dm2-effective-NPZ} for the analysis. 
Alas, the Daya Bay group~\cite{DayaBay:2018yms}, in fact, uses the same survival probability as in eq.~\eqref{P-aa-NPZ}, but with a different expression of $\Delta m^2_{ \text{effective} }$:\footnote{
In the reactor antineutrino experiments in vacuum we deal with the case of $\nu_{\alpha} = \nu_{e}$. In the symmetry discussion which follows, we will treat the generic case of $\nu_{\alpha}$ with $r_{i} (\alpha)$, see eq.~\eqref{r1-r2-def}, to show that the equivalence between the two expressions of $\Delta m^2_{ \text{effective} }$ holds also in the generic case of $\alpha = e, \mu, \tau$. } 
\begin{eqnarray} 
\Delta m^2_{ \text{effective} } \vert_{ \text{DB} } 
= 
\Delta m^2_{32} 
+ \frac{ 2 E }{ L } 
\arctan \left\{ \frac{ r_{1} \sin 2\Delta_{21}}{ r_{2} + r_{1} \cos2\Delta_{21}} \right\}. 
\label{Dm2-DayaBay}
\end{eqnarray}
Here we note that the Daya Bay group does not give the explicit form of $\Delta m^2_{ \text{effective} }$ in ref.~\cite{DayaBay:2018yms}. The expression given in eq.~\eqref{Dm2-DayaBay} is taken from refs.~\cite{Parke:2019mhz} and~\cite{DayaBay:2019bsz}. 

Let us try to understand this problem first. 

\subsection{Equivalence between Daya Bay's $\Delta m^2_{ \text{effective} }$ and ours} 
\label{sec:Dm2-DB-ours}

In fact, one can show quite easily that the Daya Bay's expression of $\Delta m^2_{ \text{effective} }$ in eq.~\eqref{Dm2-DayaBay} is identical with ours, eq.~\eqref{Dm2-effective-NPZ}. There are two ways to reveal this equivalence, one by a simple algebra with use of the formula in eq.~\eqref{math-1}, see Appendix~\ref{sec:algebraic-proof}, and the other by the reparametrization symmetry in neutrino oscillations in vacuum~\cite{Parke:2018shx,Minakata:2022zua}.\footnote{
A heuristic discussion of the reparametrization symmetry in vacuum has been initiated in ref.~\cite{Parke:2018shx}. Recently, all the eight vacuum symmetries of this type, IA-, IB-, $\cdot \cdot \cdot$, IVB-vacuum, are uncovered in ref.~\cite{Minakata:2022zua}. For the recent discussions of reparametrization symmetry in matter, see refs.~\cite{Minakata:2022yvs,Minakata:2023ice} and the references cited therein. }
Here, we present the second way, the simpler one. 

What matters in our case is the one called Symmetry IA-vacuum in the terminology of ref.~\cite{Minakata:2022zua}: 
\begin{eqnarray} 
&& 
\text{Symmetry IA-vacuum:}
\hspace{6mm}
m^2_{1} \leftrightarrow m^2_{2}, 
\hspace{8mm}
c_{12} \rightarrow - s_{12},
\hspace{8mm}
s_{12} \rightarrow c_{12},
\label{Symmetry-IA-vacuum}
\end{eqnarray} 
which implies $\Delta m^2_{21} \rightarrow - \Delta m^2_{21}$, $\Delta m^2_{31} \rightarrow \Delta m^2_{32}$, and $r_{1} \leftrightarrow r_{2}$. Then, one can show quite easily that by Symmetry IA-vacuum transformations, 
$\Delta m^2_{ \text{effective} } \vert_{ \text{DB} }$ in eq.~\eqref{Dm2-DayaBay} transforms to $\Delta m^2_{ \text{effective} }$ in eq.~\eqref{Dm2-effective-NPZ}, and vice versa. 
Thus, we have shown that $\Delta m^2_{ \text{effective} } \vert_{ \text{DB} }$ is identical with $\Delta m^2_{ \text{effective} }$ in eq.~\eqref{Dm2-effective-NPZ}, and the Daya Bay analysis~\cite{DayaBay:2018yms} uses the probability equivalent to eq.~\eqref{P-aa-NPZ}. 

\subsection{Alternative choice: $\Delta m^2_{\alpha \alpha}$, the $L/E$-independent part of $\Delta m^2_{ \text{effective} }$} 
\label{sec:Dm2-ee}

Another reactor experiment RENO~\cite{RENO:2020dxd} uses $\Delta m^2_{ \text{ee} }$ in eq.~\eqref{Dm2aa-def}, the $L/E$-independent part of $\Delta m^2_{ \text{effective} }$, in their analysis. At the first minimum of $P (\nu_{e} \rightarrow \nu_{e})$ in the atmospheric-scale oscillation, $\Delta_{31} \sim 1$ and therefore the leading-order correction to $\Delta m^2_{ \text{ee} }$ (see eq.~\eqref{Dm2-expanded}) is of order $(\Delta_{21})^3 \sim (0.03)^3 \approx 2.7 \times 10^{-5}$, which is negligibly small. Therefore, use of which $\Delta m^2_{ \text{effective} }$, Daya-Bay (= ours) or RENO, may not be so important numerically. 

\subsection{A dispute about $\Delta m^2_{ \text{effective} }$ } 
\label{sec:dispute}

Despite that there is little room for arguing which one is better on numerical ground, there is a dispute on what form of $\Delta m^2_{ \text{effective} }$ should be used. 
The authors of ref.~\cite{Parke:2019mhz} denote $\Delta m^2_{ \text{effective} }$ in eq.~\eqref{Dm2-DayaBay} as $\Delta m^2_{ \text{ee} }$(DB2), and state:~``$\Delta m^2_{ \text{ee} }$(DB2) is clearly $L/E$ (proper time of the neutrino) dependent and it is far from transparent the relationship to the fundamental neutrino parameters.'' This is somewhat unusual situation that the authors of ref.~\cite{Parke:2019mhz} recommend not to use the probability in eq.~\eqref{P-aa-NPZ} with $\Delta m^2_{ \text{effective} }$ in eq.~\eqref{Dm2-effective-NPZ} even though it was originally obtained by NPZ~\cite{Nunokawa:2005nx}. 

If this is not a simple error, some conceptual issue must be involved. In fact, in Conclusion section of ref.~\cite{Parke:2016joa}, the author lists the favorable properties of $\Delta m^2_{\alpha \alpha}$, their proposal for the effective atmospheric $\Delta m^2$, as follows: (i) It is independent of $L/E$ for all values of $L/E$, (ii) It is a simple combination of fundamental parameters. 
Then, a very likely interpretation of the above statement in ref.~\cite{Parke:2019mhz} is that the authors make the declaration of their viewpoint that {\em not} the whole $\Delta m^2_{ \text{effective} }$, but its $L/E$-independent part must be the effective atmospheric $\Delta m^2$. 

If this is the case, it cannot be sustainable. The NPZ probability in~eq.~\eqref{P-aa-NPZ}, {\em by being the exact formula}, must describe the three-generation neutrino oscillation in the whole regions of $E$ and $L$. If we remove the $L/E$-dependent part from $\Delta m^2_{ \text{effective} }$, the formula ceases to be valid in most of the available kinematical regions, including the one $\Delta_{21} \sim \mathcal{O} (1)$. All the intriguing discussions in section~\ref{sec:picture-change}, which are originally from ref.~\cite{Minakata:2007tn}, would be gone. Therefore, their assertion of $\Delta m^2_{\alpha \alpha}$ as the effective atmospheric $\Delta m^2$ must apply only to a  limited kinematical region $\Delta_{21} \ll 1$, including the one for the Daya Bay analysis. 
As the removal of the $L/E$-dependent terms from $\Delta m^2_{ \text{effective} }$ is done on an ad hoc basis in a selected kinematical region, the left-over constant $\Delta m^2_{\alpha \alpha}$ bears little resemblance to the fundamental constants. 

\subsection{Highlighting our position in the light of the debates}
\label{sec:our-position} 

In this paper, on the contrary to the viewpoint discussed in the previous section~\ref{sec:dispute}, we remain in our standpoint that $\Delta m^2_{ \text{effective} }$ in eq.~\eqref{Dm2-effective-NPZ}, or $( 2 \Delta_{\alpha \alpha} \pm \phi_{\odot} )$ in eq.~\eqref{P-aa-MNPZ}, is a physically sensible quantity {\em because} they are $L/E$ dependent. See section~\ref{sec:L-E-dep-meaning}, and the foregoing parts of this section. 
In a nutshell, the oscillation factor decomposes into the atmospheric-scale $L/E$-independent $2 \Delta_{\alpha \alpha}$, and $L/E$ dependent solar-scale $\phi_{\odot}$, owing to nature's design of the three-generation neutrinos. They jointly constitute the exact oscillation probability. 

The oscillation factor $( 2 \Delta_{\alpha \alpha} \pm \phi_{\odot} )$ may suggest that the separation between $\Delta_{\alpha \alpha}$ and $\phi_{\odot}$ terms could be done with the $\pm$ sign determining the neutrino mass ordering. 
However, we note that the mass ordering reversal is {\em not} the symmetry of our world, apparently not giving a legitimate reasoning for separating out the constant $\Delta m^2_{\alpha \alpha}$ from $\Delta m^2_{ \text{effective} }$. 

\section{Issue of notation: $\Delta m^2_{ \text{ren} }$ vs. $\Delta m^2_{ \text{ee} }$} 
\label{sec:notation}

We want to address the problem of notation in a perturbative framework called the ``renormalized helio-perturbation theory''~\cite{Minakata:2015gra}, to keep the promise of clarification in this reference. The framework allows a compact analytic description of neutrino flavor transformation around the atmospheric resonance~\cite{Wolfenstein:1977ue,Mikheyev:1985zog}. It follows several early attempts to construct the perturbation theory of neutrino oscillation in matter with the unique expansion parameter $\Delta m^2_{21} / \Delta m^2_{31}$~\cite{Arafune:1996bt,Cervera:2000kp,Freund:2001pn,Akhmedov:2004ny}. 
To simplify the formalism we have defined $\Delta m^2_{ \text{ren} } \equiv \Delta m^2_{31} - s^2_{12} \Delta m^2_{21}$, with subscript ``ren'' in shorthand for ``renormalized'', by which the Hamiltonian in the tilde basis $\tilde{H} = U_{23}^{\dagger} H_{\text{\tiny ATM} } U_{23}$ becomes\footnote{
In ref.~\cite{Minakata:2015gra} and in this section, we use the flavor-basis Hamiltonian and the $U$ matrix in the ATM convention~\cite{Martinez-Soler:2018lcy}, $U_{\text{\tiny ATM} } = \text{diag} [ 1, 1, e^{ - i \delta} ] U_{\text{\tiny PDG} }~\text{diag} [ 1, 1, e^{ i \delta} ]$ where $U_{\text{\tiny PDG} }$ is the $U$ matrix in the PDG convention~\cite{ParticleDataGroup:2026mpi}. The flavor basis Hamiltonian is given by $H_{\text{\tiny ATM} } = \text{diag} [ 1, 1, e^{ - i \delta} ] H~\text{diag} [ 1, 1, e^{ i \delta} ]$. }
\begin{eqnarray} 
\tilde{H} 
&=& 
\frac{ 1 }{ 2E } 
\left[
\begin{array}{ccc}
\lambda_{a} & 0 & c_{13} s_{13} \Delta m^2_{ \text{ren} } \\
0 & \lambda_{b} & 0 \\
c_{13} s_{13} \Delta m^2_{ \text{ren} }  & 0 & \lambda_{c} 
\end{array}
\right] 
+ 
\epsilon c_{12} s_{12} \frac{ \Delta m^2_{ \text{ren} } }{ 2 E }
\left[
\begin{array}{ccc}
0 & c_{13} & 0 \\
c_{13} & 0 & - s_{13}  \\
0 & - s_{13} & 0 
\end{array}
\right], 
\label{tilde-H}
\end{eqnarray} 
where $\epsilon \equiv \Delta m^2_{21} / \Delta m^2_{ \text{ren} }$. $\lambda_{a}$, $\lambda_{b}$, $\lambda_{c}$ denote the asymptotic eigenvalues, see ref.~\cite{Denton:2016wmg}. As the expression of $\Delta m^2_{ \text{ren} }$ is identical with $\Delta m^2_{ \text{ee} }$ (see eq.~\eqref{Dm2aa-def}), the two authors of ref.~\cite{Minakata:2015gra} have been debating on which notation should be adopted in the Hamiltonian. 

First of all, we do not know by which logic the $L/E$ independent part of $\Delta m^2_{ \text{effective} }$ must show up in the Hamiltonian. We must somehow truncate the perturbative series of $\Delta m^2_{ \text{effective} }$,  eq.~\eqref{Dm2-expanded}, to obtain $\Delta m^2_{ \text{ee} }$. But, such truncation of the $L/E$ dependent terms is justified only in the region $\Delta_{21} \ll 1$. Notice that this criterion is set merely on the ground of numerical accuracy, no basis in symmetry reasons or others. In view of Fig.~1 in ref.~\cite{Minakata:2015gra}, the perturbation theory has validity even to $L/E = 2000$ km/GeV, which amount to $\Delta_{21} = 0.19$, around where the $\Delta_{21}$ expansion starts to become inaccurate. The third order correction is smaller only by about 0.7\% than the leading term, which is comparable with the measurement accuracy~\cite{JUNO:2026tap}. Then, $\Delta m^2_{ \text{ren} } \equiv \Delta m^2_{31} - s^2_{12} \Delta m^2_{21}$, a simple combination of the physical parameters, may be a cleaner notation. 

In the case of DMP perturbation theory~\cite{Denton:2016wmg} the situation is much clearer. The region of validity of this perturbation theory extends to the solar-resonance region $\Delta_{21} \sim 1$, see ref.~\cite{Minakata:2020oxb} for a limiting procedure to this region. Therefore, usage of $\Delta m^2_{ \text{ee} }$ is strongly discouraged because $\Delta m^2_{ \text{ee} }$ is not a good approximation to $\Delta m^2_{ \text{effective} }$ at $\Delta_{21} \sim 1$. 
Notice that $\Delta m^2_{\alpha \alpha}$ is defined, to the best of our knowledges, only by the $L/E$ independent part of $\Delta m^2_{ \text{effective} }$, and there is no a priori reason to believe it as a fundamental constant. 

The present author regrets that he agreed to use $\Delta m^2_{ \text{ee} }$ in notation in ref.~\cite{Denton:2016wmg} due to insufficient understanding of the point discussed in this paper. In fact, there exists another conceptual problem. Why does only $\Delta m^2_{ \text{ee} }$ appear in the Hamiltonian, neither $\Delta m^2_{\mu \mu}$ nor $\Delta m^2_{\tau \tau}$? We recall that the Hamiltonian serves not only for $\nu_{e}$-related channels but also for $\nu_{\mu}$- and $\nu_{\tau}$-related channels. We feel strange to stick to use only $\Delta m^2_{ \text{ee} }$ in the element of the Hamiltonian. Thus, we recommend using the plain notation such as $\Delta m^2_{ \text{ren} }$.

\section{Effective two-frequency description and $\Delta m^2_{ \text{effective} }$ in matter}
\label{sec:Dm2-eff-matter}

In this section, we give a short discussions of the two-frequency description and the effective atmospheric $\Delta m^2$ in matter. We restrict our discussion into the constant density matter. 

\subsection{Matter-MNPZ probability and $\Delta m^2_{ \text{effective} }$}
\label{sec:MNPZ-matter}

We repeat an analogous treatment in vacuum as given in section~\ref{sec:MNPZ-P} to derive the MNPZ formula in matter. One can introduce the $\widetilde{U}$ matrix analogous to the $U$ matrix in vacuum, to diagonalize the flavor basis Hamiltonian $H$ in matter as 
\begin{eqnarray} 
&&
\frac{1}{2E}
\widetilde{U}^{\dagger}
\left\{ 
U \left[
\begin{array}{ccc}
m^2_{1} & 0 & 0 \\
0 & m^2_{2} & 0 \\
0 & 0 & m^2_{3} \\
\end{array}
\right] U^{\dagger}
+
\left[
\begin{array}{ccc}
a(x) & 0 & 0 \\
0 & 0 & 0 \\
0 & 0 & 0
\end{array}
\right] 
\right\} \widetilde{U} 
= 
\frac{1}{2E} 
\left[
\begin{array}{ccc}
\lambda_{1} & 0 & 0 \\
0 & \lambda_{2} & 0 \\
0 & 0 & \lambda_{3} \\ 
\end{array}
\right] 
\label{matter-H}
\end{eqnarray}
where $a(x) = 2 \sqrt{2} G_F N_e E$ in eq.~\eqref{matter-H} denote the Wolfenstein matter potential~\cite{Wolfenstein:1977ue}.\footnote{
In eq.~\eqref{matter-H} , $G_F$ is the Fermi constant. $N_e$ and $N_n$ are the electron and neutron number densities in matter. $\rho$ and $Y_e$ denote, respectively, the matter density and number of electrons per nucleon in matter. 
$a(x) = 1.52 \times 10^{-4} \left( \frac{Y_e \rho}{\rm g\,cm^{-3}} \right) \left( \frac{E}{\rm GeV} \right) {\rm eV}^2$. } 
$\nu_{\alpha} = U_{\alpha i} \nu_{i}$ in vacuum, and $\nu_{\alpha} = \widetilde{U}_{\alpha i} \widetilde{\nu}_{i}$ in matter, where $\nu_{i}$ and $\widetilde{\nu}_{i}$ are mass eigenstate basis in vacuum and in matter, respectively.  

We can parametrize the $\widetilde{U}$ matrix in the same way as the $U$ matrix in vacuum~\cite{ParticleDataGroup:2026mpi}, by replacing the vacuum mixing angles $\theta_{ij}$ $(ij=1,2,3)$ to the matter mixing angles $\widetilde{\theta}_{ij}$, and the lepton KM phase $\delta$~\cite{Kobayashi:1973fv} to $\widetilde{\delta}$. The exact expressions of the eigenvalues $\lambda_{i}$, the matter mixing angles $\widetilde{\theta}_{ij}$ and $\widetilde{\delta}$ are obtained in ref.~\cite{Zaglauer:1988gz}. 
Then, the the disappearance probability $P (\nu_{\alpha} \rightarrow \nu_{\alpha})$ ($\alpha = e, \mu, \tau$) in matter can be written by the same form as in eq.~\eqref{Paa-vac}, but with the replacements 
\begin{eqnarray} 
&&
U_{\alpha i} (\theta_{12}, \theta_{13}, \theta_{23}, \delta )
\rightarrow 
\widetilde{U}_{\alpha i} 
( \widetilde{\theta}_{12}, \widetilde{\theta}_{13}, \widetilde{\theta}_{23}, \widetilde{\delta} ),
\nonumber\\
&& 
\Delta_{ji} \rightarrow \widetilde{\Delta}_{ji} \equiv 
( \lambda_{j} - \lambda_{i} ) L / (4E). 
\label{replacement} 
\end{eqnarray}
In matter, the antineutrino probability is given by $P (\bar{\nu}_{\alpha} \rightarrow \bar{\nu}_{\alpha}: a, \widetilde{\delta}) = P (\nu_{\alpha} \rightarrow \nu_{\alpha}: - a, - \widetilde{\delta})$. 
Then, the matter-MNPZ probability formula can be obtained in the same form as in vacuum, eq.~\eqref{P-aa-MNPZ}, by the same replacements in eq.~\eqref{replacement}. $\widetilde{r}_{i} (\alpha)$ ($i=1,2$) can be obtained by using eq.~\eqref{r1-r2-def} but with the replacements $U_{\alpha i} \rightarrow \widetilde{U}_{\alpha i}$. $\widetilde{\Delta}_{\alpha \alpha}$ and $\widetilde{\phi}_{\odot}$ can also be obtained by the similar manner. 

The same exercise as done in vacuum in section~\ref{sec:NPZ-P} can be set up in matter by minimizing the solar-scale terms to obtain the two-frequency description of the disappearance oscillation probability. Then, we should obtain the matter-NPZ formula, corresponding to the vacuum version in eq.~\eqref{P-aa-NPZ}. The expression of $\Delta m^2_{ \text{effective} }$ in matter is given by 
\begin{eqnarray} 
&&
\Delta m^2_{ \text{effective} } \vert_{ \text{matter} }
= 
( \lambda_{3} - \lambda_{1} ) 
- 
\frac{ 2 E }{ L }
\arctan \left\{ \frac{ \widetilde{r}_{2} \sin 2 \widetilde{\Delta}_{21} }{ \widetilde{r}_{1} + \widetilde{r}_{2} \cos 2 \widetilde{\Delta}_{21} } \right\}. 
\label{Dm2-effective-NPZ-matter} 
\end{eqnarray}
By appealing the same algebra as used in section~\ref{sec:equivalence}, one must be able to show that \\ 
$\Delta m^2_{ \text{effective} } \vert_{ \text{matter} } L / (2E) = 2 \widetilde{\Delta}_{\alpha \alpha} \pm \widetilde{\phi}_{\odot}$. 
Since our treatment is based on the exact solution of the three-neutrino oscillation in uniform-density matter, see refs.~\cite{Zaglauer:1988gz,Kimura:2002wd}, we assume that the matter-NPZ (and MNPZ) formula and $\Delta m^2_{ \text{effective} }$ in matter in eq.~\eqref{Dm2-effective-NPZ-matter} are exact. 

\subsection{Dynamical property of $\Delta m^2_{ \text{effective} }$ in matter} 
\label{sec:dynamical-Dm2}

Now, we observe that $\Delta m^2_{ \text{effective} }$ in matter becomes even more dynamical than that in vacuum. That is, $\Delta m^2_{ \text{effective} }$ and the probability start to have dependencies other than $L/E$ due to the introduction of the mass-dimension two quantity $a = 2 \sqrt{2} G_F N_e E$, the Wolfenstein matter potential. $a$ can appear directly $\Delta m^2_{ \text{effective} }\vert_{ \text{matter} }$, and through the new dimensionless quantities such as $a / \Delta m^2_{31} \propto E$ and $\Delta_{a} \equiv a L / 4E \propto L$. 

Let us display some of these features by an approximate calculation of $\Delta m^2_{ \text{effective} }\vert_{ \text{matter} }$. We use the expanded form given in eq.~\eqref{Dm2-expanded}, and convert all the quantities to the ones in matter, $\Delta m^2_{31} \rightarrow \lambda_{3} - \lambda_{1}$, $r_{2} \rightarrow \widetilde{r}_{2}$, etc. To first order we obtain 
\begin{eqnarray} 
&&
\Delta m^2_{ \text{effective} } (\alpha) \vert_{ \text{matter} } 
=
( \lambda_{3} - \lambda_{1} ) - \widetilde{r}_{2} (\alpha) ( \lambda_{2} - \lambda_{1} ). 
\label{Dm2-matter-1st}
\end{eqnarray}
Notice that the smallness of the higher order terms of $\widetilde{\Delta}_{21}$ is not guaranteed in general, because $\widetilde{\Delta}_{21}$ can become $\sim \mathcal{O} (1)$ in some kinematical region such as the solar resonance. 

Here, we just want to show appearance of the $a$-dependent terms in $\Delta m^2_{ \text{effective} } (\alpha) \vert_{ \text{matter} }$ in eq.~\eqref{Dm2-matter-1st}. For this purpose we use the helio-perturbation theory~\cite{Minakata:2015gra} mentioned in section~\ref{sec:notation} to estimate the right-hand side of eq.~\eqref{Dm2-matter-1st}. We examine the $\nu_{\mu}$ channel and take the NMO for simplicity. We remind that this framework provides a valid description of $P (\nu_{\mu} \rightarrow \nu_{\mu})$ in region $| Y_e \rho E | \leq 20~\rm g~cm^{-3}~\rm GeV$ and $L/E \leq 2000~\rm km/GeV$ for the both NMO and IMO. See Fig.~1 in ref.~\cite{Minakata:2015gra}. Then, we obtain $\Delta m^2_{ \text{effective} } (\mu) \vert_{ \text{matter} }$ as 
\begin{eqnarray} 
&&
\hspace{-7mm}
\Delta m^2_{ \text{effective} } (\mu) \vert_{ \text{matter} } 
=
\Dmsqren \left( 1 - r_{A} \right) + a \frac{ 2 s^2_{13} }{ \left( 1 - r_{A} \right) } 
- \widetilde{r}_{2} (\mu) 
\left[ 
a \left\{ 1 - \frac{ 1 }{ 2 } \frac{ 2 s^2_{13} }{ \left( 1 - r_{A} \right) } \right\} 
- \cos 2\widetilde{\theta}_{12} \Delta m^2_{21} 
\right], 
\nonumber \\
\label{Dm2-matter-helioP}
\end{eqnarray}
where $r_{A} \equiv a / \Delta m^2_{ \text{ren} }$. 
To obtain eq.~\eqref{Dm2-matter-helioP}, we have stayed  in the region $r_{A} < 1$ and have used the approximation $s_{13} \ll 1$ to keep the expression simple. 
$\widetilde{r}_{2} (\mu)$ is given, with a compact notation $\widetilde{J}_{r} \equiv \widetilde{c}_{12} \widetilde{s}_{12} \widetilde{c}_{23} \widetilde{s}_{23} \widetilde{s}_{13}$, by 
\begin{eqnarray} 
&&
\widetilde{r}_{2} (\mu) 
= 
\frac{ \widetilde{c}^2_{12} \widetilde{c}^2_{23} + \widetilde{s}^2_{12} \widetilde{s}^2_{23} \widetilde{s}^2_{13} - 2 \widetilde{J}_{r} \cos \widetilde{\delta} }{ ( \widetilde{c}^2_{23} + \widetilde{s}^2_{23} \widetilde{s}^2_{13} ) }. 
\label{r2-mu}
\end{eqnarray}
In the framework of ref.~\cite{Minakata:2015gra} only $\theta_{13}$ becomes matter-modified, $\widetilde{\theta}_{13} = \phi$, leaving $\widetilde{\theta}_{12}$ and $\widetilde{\theta}_{23}$ at their vacuum values. In the leading order in $s_{13}$, 
$\sin \phi = s_{13} / ( 1 - r_{A} )$. Therefore, $\widetilde{r}_{2} (\mu)$ is of order unity, in region of the validity of our framework. 

Thus, we have shown that $\Delta m^2_{ \text{effective} } (\mu) \vert_{ \text{matter} }$ becomes even more dynamical in matter 
by having the terms proportional to $a = 2 \sqrt{2} G_F N_e E$, and through dimensionless $a / \Delta m^2_{31}$ factor, which are proportional to $E$ and the planet matter density. It is unlikely that exact treatment based on ref.~\cite{Zaglauer:1988gz} would alter this statement. 

\subsection{Conceptual erratum to JHEP \textbf{05} (2017) 043~\cite{Minakata:2017ahk}}
\label{sec:erratum} 

In ref.~\cite{Minakata:2017ahk} the present author addressed the problem of effective two-frequency description of neutrino oscillation in matter and 
$\Delta m^2_{ \text{effective} } (\mu) \vert_{ \text{matter} }$. Using the different perturbative framework developed in ref.~\cite{Asano:2011nj}, we have observed that $\Delta m^2_{ \text{effective} } (\mu) \vert_{ \text{matter} }$ depends not only on $r_{A}$ but also on $\Delta_{a} \equiv a L / 4E \propto L$. 

It is stated in section 5 in ref.~\cite{Minakata:2017ahk} that: ``Because of the $L$-dependence of $\Delta m^2_{\mu \mu} (a)$, unfortunately, we cannot consider it as the sensible quantity as the effective parameter which describes the physics of $\nu_{\mu}$ survival probability in matter''. However, this statement cannot hold in the light of our understanding discussed throughout this paper. That is, $\Delta m^2_{ \text{effective} }$ {\em is dynamical}, $L/E$ dependent in vacuum, and has additional $L$ and $E$ dependence in matter. This section is to correct this and the similar statements made in ref.~\cite{Minakata:2017ahk} due to the author's insufficient understanding of $\Delta m^2_{ \text{effective} }$. 

Other than this conceptual issue, no error is detected in the theoretical formulas and the analysis results given in this reference. In addition to the topics discussed in the present paper, the problem of two-frequency description of the {\em appearance} probability in vacuum is examined in ref.~\cite{Minakata:2017ahk}. 

\subsection{Is effective two-frequency description adequate in matter?}
\label{sec:two-frequency} 

Our treatment in section~\ref{sec:MNPZ-matter} leads us to the exact matter-MNPZ probability formula by simple mappings of the parameters from the vacuum to matter, which may be useful to discuss the small matter-effect correction in the reactor measurement. However, the question remains on how revealing the formula could be in the generic kinematical phase space for the three-generation neutrino oscillation. 

The full structure of the three-generation neutrino world involves the rich structure such as the atmospheric- and solar-scale resonances, which may be described by the exact treatment~\cite{Zaglauer:1988gz,Kimura:2002wd}, or the globally valid perturbative frameworks~\cite{Denton:2016wmg,Agarwalla:2013tza}. 
In coming close to the resonance points, the eigenvalue difference becomes smallest in the entire region of the energy $E$ for a given baseline. As a consequence, at around the solar resonance the eigenvalue splitting satisfies the relation, the atmospheric splitting $\gg$ the solar splitting, that is, $| \lambda_{3} - \lambda_{2} | \gg | \lambda_{2} - \lambda_{1} |$ (NMO), and $| \lambda_{3} - \lambda_{1} | \gg | \lambda_{2} - \lambda_{1} |$ (IMO). Whereas at around the atmospheric resonance the relation is reversed, the solar splitting becomes greater than the atmospheric splitting by a factor of several. In generic phase space, the two splittings are comparable in size, apart from the above-stated regions and the asymptotic one $|a| \gg | \Delta m^2_{31} |$. See Fig.~1 in ref.~\cite{Denton:2016wmg}. 

Under this circumstance, the effective two-frequency description of neutrino oscillation in matter may require the both ``solar'' and ``atmospheric'' $\Delta m^2_{ \text{effective} }$, but it is not obvious whether we could find the way for their proper definition. 
These cautious remarks given here may suggest that the concept of the two-frequency description and $\Delta m^2_{ \text{effective} }$ in matter should be taken with a grain of salt at this stage. 

\section{Summary and discussions} 
\label{sec:conclusion} 

We have started our discussion from the naive two-frequency description of the three-generation neutrino oscillations. It naturally led us to introduce the concept of the effective $\Delta m^2$, the single atmospheric-scale $\Delta m^2_{ \text{effective} }$, which unifies the effects of $\Delta m^2_{31}$-$\Delta m^2_{32}$ driven oscillations. In this paper, we tried to convey the readers its charm, the interesting but highly nontrivial characteristics, and the resultant controversy about this quantity. We aimed at settling all the relevant questions raised on $\Delta m^2_{ \text{effective} }$, if they are brought to our attention and as far as it is possible. 

Following the original line of approach to the effective two-frequency description of neutrino oscillations~\`a la NPZ~\cite{Nunokawa:2005nx}, we have marked a milestone by showing the exactness of $\Delta m^2_{ \text{effective} }$ and the survival probability. This is done by proving the equivalence between the NPZ and the MNPZ~\cite{Minakata:2007tn} forms of the probabilities, eq.~\eqref{P-aa-NPZ} and~eq.\eqref{P-aa-MNPZ}, whose latter is, by construction, identical to the exact oscillation probability in vacuum. It settles all possible discussions on which expression is most appropriate for $\Delta m^2_{ \text{effective} }$. 
Through the process of these discussions we illuminate how and why $\Delta m^2_{ \text{effective} }$ is $L/E$ dependent. 

We give supplementary remarks in below on some of the points we have made. An itemized several-line statements should suffice as any one of our observations is so simple. 
\begin{enumerate} 

\item 
Description of the three-generation neutrino oscillation in terms of the solar- and atmospheric-scale two frequencies necessitates the latter being $L/E$ dependent. The reason is that the constant two frequencies cannot describe the system with the three frequencies. 

\item 
The reason why $\Delta m^2_{ \text{effective} }$ is $L/E$ dependent and how the dependence enter are easy to understand. 
This is the atmospheric-scale quantity, but with the solar oscillation effect renormalized. As the solar oscillation has a different $L/E$ dependence from the atmospheric one, the $L/E$ dependence enter into $\Delta m^2_{ \text{effective} }$ via the solar variable.  

\item 
An inverse question: 
Why is the $L/E$ dependent $\Delta m^2_{ \text{effective} }$ sufficient to describe {\em exactly} the three-generation world? The answer is provided by the MNPZ formula. The exact oscillation probability is reproduced because of the $L/E$ dependent solar effect renormalization in $\Delta m^2_{ \text{effective} }$.  

\item 
We have confirmed the dynamical feature of $\Delta m^2_{ \text{effective} }$, 
the $L/E$ dependences and an individual $E$- and/or $L$-dependences in matter. A hierarchy between the atmospheric $\gg$ the solar level splittings at the solar resonance, and a converse (milder) hierarchy at the atmospheric resonance poses a highly nontrivial issue for the effective two-frequency description in matter. 

\end{enumerate}

We remark that the exactness of the probability formula and 
$\Delta m^2_{ \text{effective} }$ has implications: The formula must be valid in the whole regions of $E$ and $L$. Suppose that some authors prefer $\Delta m^2_{ \text{effective} }$ be $L/E$ independent. The MNPZ decomposition $\Delta m^2_{ \text{effective} } = \Delta m^2_{\alpha \alpha} \pm (2E/L) \phi_{\odot}$ with $\phi_{\odot}$ as the purely solar-scale variable, may suggest that $\Delta m^2_{\alpha \alpha}$, which is composed of the mixing parameters only, qualifies as an effective atmospheric $\Delta m^2$. Such consideration might have led some authors to propose $\Delta m^2_{ \text{effective} } = \Delta m^2_{ \text{ee} }$ ansatz, as mentioned in section~\ref{sec:dispute}. 

However, if $\Delta m^2_{\alpha \alpha}$ alone drive the $\Delta m^2_{ \text{effective} }$ oscillation without $\phi_{\odot}$, of course, it fails to describe the JUNO region. This means that the applicability of the $\Delta m^2_{ \text{effective} } = \Delta m^2_{\alpha \alpha}$ ansatz must be restricted, practically, to the region $\Delta_{21} \ll 1$. Then, one must pay the price. Even though a constant $\Delta m^2_{\alpha \alpha}$ is isolated from the $L/E$ dependent forest, it is hard to believe that this is a fundamental constant because of the ad hoc procedure to modify the exact probability formula. This feature discourages $\Delta m^2_{ \text{ee} }$ from appearing in the Hamiltonian, as argued in section~\ref{sec:notation}. It appears also that there is no symmetry by which one can assign the different transformation properties to $\Delta m^2_{\alpha \alpha}$ and $\phi_{\odot}$ to single out the former. 

Recently, JUNO reported their 207 days data of reactor antineutrino measurement for the atmospheric mass-squared splitting $\Delta m^2_{31}$, reaching $\simeq 1\%$ precision for the first time~\cite{JUNO:2026tap}. Then, it is natural to see that the analyses for the mass-ordering determination starts to shift toward the ``global analysis'' type. They include: JUNO combined with the atmospheric~\cite{Super-Kamiokande:2023ahc,IceCubeCollaboration:2024ssx,KM3NeT:2024ecf} and/or the accelerator~\cite{T2K:2025yoy,NOvA:2025tmb} neutrino experiments, or the NuFit results~\cite{Esteban:2024eli} as the external inputs~\cite{JUNO:2026tap}. Among them, the atmospheric neutrino constraints are so powerful to result in the higher-than~$3 \sigma$ preference of the normal mass ordering over the inverted, see e.g., refs.~\cite{Jin:2026ejo,Eller:2026urd}. A possible effect of the T2K$-$NOvA tension is also addressed~\cite{Machado:2026shm}. 

Obviously the utility of $\Delta m^2_{ \text{effective} }$ comes under scrutiny in the era of the global analysis. If the analysis done by Capozzi {\it et al.}~\cite{Capozzi:2013psa} with use of the MNPZ formula has advantage over the conventional ones in some aspects, such as the separated treatment of the solar and atmospheric oscillation effects, this possibility may warrant further examinations. Eventually a super-high precision era will come. JUNO expects to achieve measurement of 0.2\% accuracy for the atmospheric $\Delta m^2_{31}$ by its six years operation~\cite{JUNO:2022mxj}. Then, implementation of the matter effect into the analysis code would become the necessity. This may be done with the MNPZ formula in matter discussed in section~\ref{sec:Dm2-eff-matter}. 

\begin{acknowledgments} 

The author thanks Stephen Parke for his great enthusiasm for the problem of $\Delta m^2_{ \text{effective} }$, his patience in encouraging and leading the discussions on this matter with the people involved, including the present author. Without such his passion, the author's understanding of the problem of $\Delta m^2_{ \text{effective} }$ had never reached the stage as that presented in this manuscript. 

\end{acknowledgments}

\appendix

\section{Two different expressions of $\phi_{\odot}$} 
\label{sec:phi-relations} 

We have defined $\phi_{\odot}$ in eq.~\eqref{sin-cos-phi-expression}, and gave another expression of $\phi_{\odot}$ in eq.~\eqref{phi-expression}. We show that they are equivalent to each other, apart from the quadrant information. We start from the latter, 
$\phi_{\odot} + ( r_{1} - r_{2} ) \Delta_{21} = \arctan \left\{ ( r_{1} - r_{2} ) \tan \Delta_{21} \right\}$, and derive the former, eq.~\eqref{sin-cos-phi-expression}. We define $x$ as 
\begin{eqnarray} 
&&
x = ( r_{1} - r_{2} ) \tan \Delta_{21} 
\label{x-def}
\end{eqnarray}
and use a formula for the trigonometric functions 
\begin{eqnarray} 
&&
\arctan x 
= 
\arcsin \frac{ x }{ \sqrt{ 1 + x^2 } } 
= 
\arccos \frac{ 1 }{ \sqrt{ 1 + x^2 } }. 
\label{formula}
\end{eqnarray}
Then, we obtain 
\begin{eqnarray} 
&&
\cos [ \phi_{\odot} + ( r_{1} - r_{2} ) \Delta_{21} ] 
= 
\frac{ 1 }{ \sqrt{ 1 + x^2 } }, 
\nonumber\\
&&
\sin [ \phi_{\odot} + ( r_{1} - r_{2} ) \Delta_{21} ] 
= 
\frac{ x }{ \sqrt{ 1 + x^2 } }, 
\label{phi-eq1}
\end{eqnarray}
which can be solved for $\cos \phi_{\odot}$ and $\sin \phi_{\odot}$ as 
\begin{eqnarray} 
&&
\cos \phi_{\odot} 
= 
\frac{ 1 }{ \sqrt{ 1 + x^2 } } 
\biggl\{ \cos [ ( r_{1} - r_{2} ) \Delta_{21} ] 
+ \sin [ ( r_{1} - r_{2} ) \Delta_{21} ] ( r_{1} - r_{2} ) \tan \Delta_{21} \biggr\}, 
\nonumber \\
&& 
\sin \phi_{\odot} 
= 
\frac{ 1 }{ \sqrt{ 1 + x^2 } } 
\biggl\{ \cos [ ( r_{1} - r_{2} ) \Delta_{21} ] ( r_{1} - r_{2} ) \tan \Delta_{21} - \sin [ ( r_{1} - r_{2} ) \Delta_{21} ] \biggr\}. 
\label{phi-eq2}
\end{eqnarray}
Using the formula which is easily obtainable 
\begin{eqnarray} 
&& 
\frac{1}{ \sqrt{ 1 + x^2 } } 
= 
\frac{ \cos \Delta_{21} }{ \sqrt{ 1 - 4 r_{1} r_{2} \sin^2 \Delta_{21} } }
\label{phi-eq3}
\end{eqnarray}
we reproduce eq.~\eqref{sin-cos-phi-expression} via a little algebras with inserting $1 = ( r_{1} + r_{2} )$ into appropriate places. Therefore, eq.~\eqref{phi-expression} and eq.~\eqref{sin-cos-phi-expression} are the two equivalent expressions of $\phi_{\odot}$. 

\section{Identity between the Daya Bay and NPZ expressions of $\Delta m^2_{ \text{effective} }$}
\label{sec:algebraic-proof}

The identity between the Daya Bay and NPZ expressions of $\Delta m^2_{ \text{effective} }$, eqs.~\eqref{Dm2-DayaBay} and~\eqref{Dm2-effective-NPZ}, can be written as 
\begin{eqnarray} 
&& 
\arctan \left\{ \frac{ \sin D }{ \frac{r_1}{r_2} + \cos D } \right\} 
+ 
\arctan \left\{ \frac{ \sin D }{ \frac{r_2}{r_1} + \cos D } \right\} 
= D
\label{NPZ-DB2}
\end{eqnarray}
where $D \equiv 2\Delta_{21} = \frac{ \Delta m^2_{21} L}{ 2E }$. We use the formula in eq.~\eqref{math-1}, the upper sign, and insert 
\begin{eqnarray} 
&& 
x = \frac{ \sin D }{ \frac{r_1}{r_2} + \cos D }, 
\hspace{12mm}
y = \frac{ \sin D }{ \frac{r_2}{r_1} + \cos D }, 
\label{x-y-def}
\end{eqnarray}
into the left-hand side. With $x$ and $y$ in eq.~\eqref{x-y-def} one can show easily that 
\begin{eqnarray} 
&& 
\frac{ x+y} {1 - x y} = \tan D. 
\label{identity-RHS}
\end{eqnarray}
That is, the right-hand side of eq.~\eqref{math-1} is given by $D$. Equation~\eqref{NPZ-DB2} holds, which means that the Daya Bay and NPZ expressions of $\Delta m^2_{ \text{effective} }$ is equivalent to each other.

\end{document}